\documentclass[twocolumn,showpacs,preprintnumbers,amsmath,amssymb]{revtex4-1}

\usepackage{graphicx}% Include figure files
\usepackage{dcolumn}% Align table columns on decimal point
\usepackage{bm}% bold math

\def\Tc{$T_{c}$}
\def\Tchimax{$T^{\chi}_{max}$}
\def\TchimaxN{$T^{\chi}_{max}(N)$}

\def\nup{$\nu'$}
\def\cio{$\chi_{0}$}
\def\ket#1{\left\vert #1 \right\rangle}
\def\Tunit{$J/k_{B}$}
\def\mo{$m_{0}$}
\def\betnu{$\beta/\nu'$}
\def\gamnu{$\gamma/\nu'$}

\begin{document}

\preprint{APS/123-QED}

\title{Analytical solution of thermal magnetization on memory stabilizer structures}

\author{Yu Tomita}
\author{C. Ricardo Viteri}
\author{Kenneth R. Brown}
\altaffiliation{Author to whom correspondence should be addressed. Electronic mail: ken.brown@chemistry.gatech.edu}
\affiliation{Schools of Chemistry and Biochemistry; Computational Science and Engineering; and Physics, Georgia Institute of Technology, Atlanta, Georgia 30332, USA}

\date{\today}% It is always \today, today,
% but any date may be explicitly specified
\begin{abstract}
We return to the question of how the choice of stabilizer generators affects the preservation of information on structures whose degenerate ground state encodes a classical redundancy code. Controlled-not gates are used to transform the stabilizer Hamiltonian into a Hamiltonian consisting of uncoupled single spins and/or pairs of spins. This transformation allows us to obtain an analytical partition function and derive closed form equations for the relative magnetization and susceptibility. These equations are in agreement with the numerical results presented in [Phys. Rev. A \textbf{80}, 042313 (2009)] for finite size systems. Analytical solutions show that there is no finite critical temperature, \Tc$=0$, for all of the memory structures in the thermodynamic limit. This is in contrast to the previously predicted finite critical temperatures based on extrapolation. The mismatch is a result of the infinite system being a poor approximation even for astronomically large finite size systems, where spontaneous magnetization still arises below an apparent finite critical temperature. We extend our analysis to the canonical stabilizer Hamiltonian. Interestingly, Hamiltonians with two-body interactions have a higher apparent critical temperature than the many-body Hamiltonian.
\end{abstract}

\pacs{03.67.Lx, 03.67.Pp, 64.60.an, 64.60.De, 75.10.Hk, 75.10.Pq, 75.40.Mg, 75.40.Cx, 89.75.Da}% PACS, the Physics and Astronomy
\maketitle

\section{\label{sec:Int}Introduction}
The equivalent of a magnetic memory for quantum information would consist of a macroscopic number of qubits with multi-qubit interactions that create a single stable qubit memory. The free energy of the system would depend upon an external control to spontaneously break global symmetry in the presence of environment-induced fluctuations. Kitaev's toric code in a four dimensional lattice would achieve this task~\cite{Kitaev2003, Dennis2002}, but its implementation seems currently unlikely. Bravyi and Terhal have recently shown that a two-dimensional self-correcting quantum memory may not exist~\cite{Bravyi2009}. If dimensionality is an engineering limitation, the solution may be self-correcting memories of finite size based on concatenated codes in which the number of qubits involved in each interaction grows with the lattice size~\cite{Bacon2008}. The classical concatenated triple modular redundancy code in the formalism of quantum stabilizers using the standard choice of generators fulfills this prerequisite for classical memory.

The stabilizer for a subspace is defined as the group of Pauli operators that act trivially on a code space and whose eigenvalues are +1. The code space is the degenerate ground state of a Hamiltonian built from the stabilizer elements with negative couplings. The triple-modular redundancy code is a textbook example for introducing the idea of stabilizer error correcting codes~\cite{Nielsen_book}. Classical error correcting codes represent a subset of quantum error correcting codes that only protect against classical bit-flip errors but not phase errors~\cite{Viteri2009}. At each level of concatenation $k$, the logical bit consists of three bits of level $k-1$, and correction works by majority vote at the lowest level first and then working up. The $k$-th level of concatenated code contains $3^k$ bits or classical spins, and it can always correct a maximum of $2^{k}-1$ errors on the physical bits. The increase of $k$ leads to many-body operators that test the parity of $\frac{2}{3} \times 3^k$ bits at once. This exponential increase in the many-body nature of the Hamiltonian makes the physical construction of such a system unrealistic.

\begin{figure}
\centering
\includegraphics[width=0.45\textwidth]{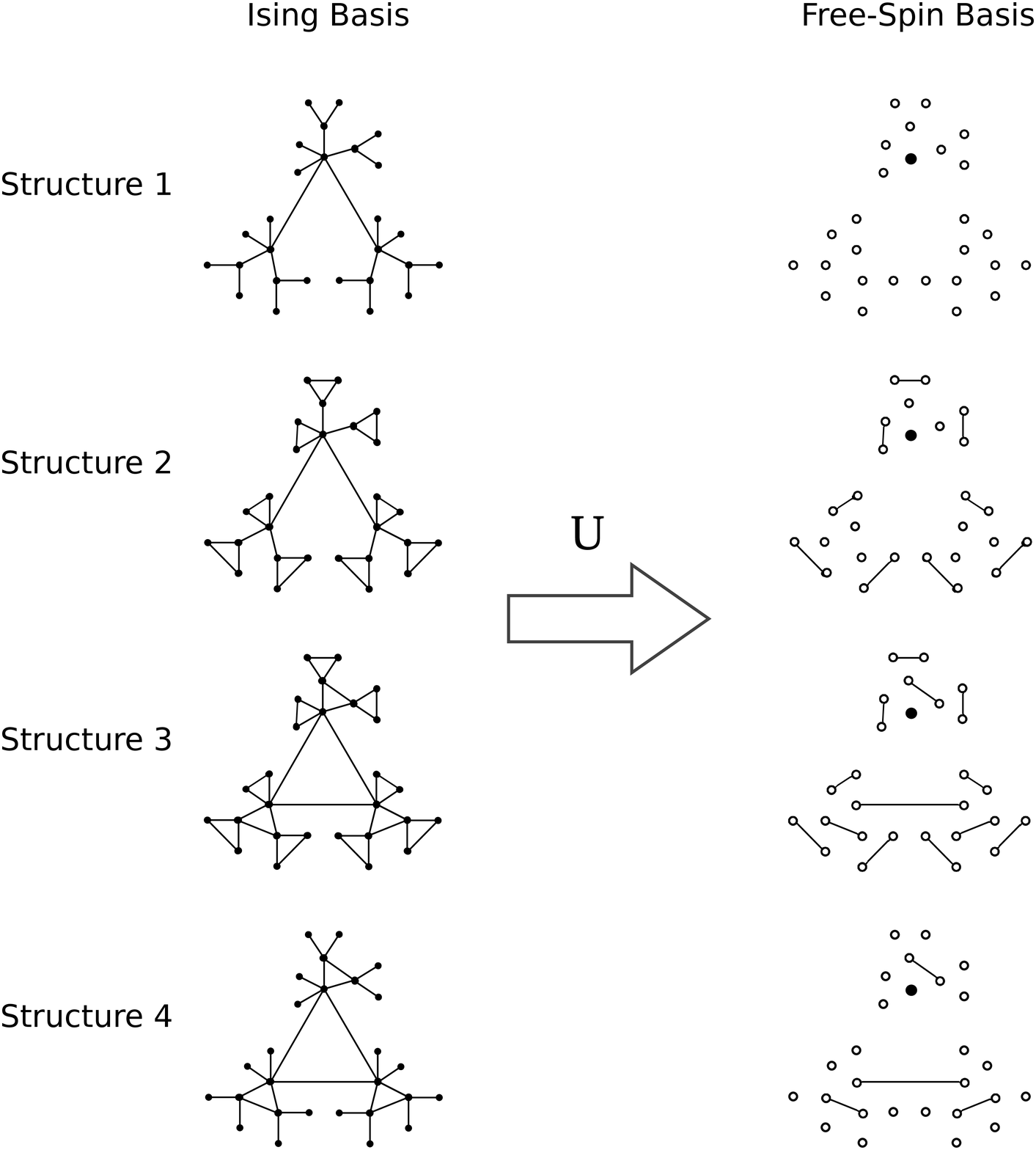}
\caption{\label{fig:structures} Transformation of the memory stabilizer structures generated by two body interactions from the Ising basis to the free-spin basis. Black dots and open circles are spin sites (qubits), and the lines show pairs of interacting spins (generators). In the free-spin basis, the black dot without interactions is a single free spin, open circles are independent spins in a magnetic field, and the connected circles are independent pairs of interacting spins in a magnetic field. The interaction strength $J$ is constant (see Eq.~\ref{eq:IsingHamiltonian}). The total number of bits increases with concatenation level, $k$, as $3^{k}$. Only $k=3$ level structures are shown.}
\end{figure}

An alternative choice uses only elements that test a pairwise agreement. This set of Pauli operators generates the same stabilizer group and represent an Ising Hamiltonian with characteristic thermodynamic and kinetic properties. Using Monte Carlo simulations, we examined the thermal magnetization of this pairwise choice of stabilizers (Structure 1 in Fig.~\ref{fig:structures}) and the effect of adding non-independent stabilizers to the Hamiltonian (Structures 2 and 3)~\cite{Viteri2009}. For Structure 1, $3^k-1$ independent stabilizer elements form a tree. Structures 2 and 3 are modifications that include cycles in the structure. The cycles are equivalent to choosing an overcomplete set of stabilizer generators.

In this paper, we analytically evaluate the choice of stabilizer generators on the preservation of information . Specifically, a unitary operator is constructed from controlled-not gates that converts a Hamiltonian representing an Ising tree into a Hamiltonian of uncoupled spins in a magnetic field. Applying the same unitary operator to the tree-like graphs of Structures 2, 3, and 4 yields partition functions corresponding to a collection of independent single spins and independent pairs of spins. A slight modification of the sequence allows us to calculate the analytical magnetization of the canonical stabilizer Hamiltonian. The results presented here agree with our previous numerical work for relatively small, finite-size systems. Closed form partition functions for each of the four self-correcting memory structures allow us to examine the problem at much larger $k$.

A direct measurement of the degree of preservation of the information can be read from the spontaneous magnetization at zero magnetic field. Below a certain temperature, a single spin, $s_0$, is sufficient to bias the system into one of the two states of broken symmetry. The finite-size system develops spontaneous magnetization and the single order parameter $m_0=(\sum_{j=0}^{N-1}\left<s_0s_j\right>)/N$ approaches the value of 1~\cite{Chandler_book}. The stability of the structure, as measured by the temperature range in which \mo\ is preserved, depends on the energy barrier that separates the two ground states and the number of pathways that traverse the barrier.

Structure 1 is an example of an Ising tree with free boundaries. The Ising model on Cayley trees results in partition functions that are equivalent to free spins~\cite{Stosic2005,Eggarter1974}. Our previous analysis, based on $N = 81, 243, 729$, and $2187$ bits ($k=4-7$) and under the assumption that Fisher's finite-size scaling method~\cite{Fisher1972} applies to these type of Ising graphs, yielded a non-zero \Tc. But contrary to Sierpinski fractals~\cite{Carmona1998, Monceau1998, Monceau2001}, where a few data points seem to be enough to forecast \Tc\ correctly, the finite-size scaling fails to describe magnetic susceptibility peaks shifted away from \Tc=0. For Ising trees and for Sierpinski gaskets, the relative magnetization approaches zero in the thermodynamic limit~\cite{Eggarter1974, Mullerha1974, GefenSG1984, Gefen1980}, but it persists for very large systems (comparable to the number of hadrons in the universe)\cite{Liu1985, Stosic2005}. The nature of the magnetic phase transition for an infinite system is not applicable to systems of laboratory dimensions.
We find similar behavior in Structures 2, 3, and 4, but with higher apparent critical temperatures (defined as the temperature where magnetic susceptibility reaches its maximum). Surprisingly, the canonical choice of elements to generate the concatenated three-bit error-correction code exhibits the lowest of the finite-size apparent critical temperatures.

\section{\label{sec:methods}CNOT transformations and Ising systems}

The algebra of controlled-nots (CNOTs) and Pauli $Z$ operators from quantum computation is used to find analytical solutions for the internal energy and magnetization of the Ising structures in Fig. \ref{fig:structures}. Following standard notation, the spin or qubit basis is labeled $\ket{0}$ and $\ket{1}$ with the Pauli $Z$ operator in the computational basis acting as $Z\ket{x}=(-1)^x\ket{x}$, where $x$ equals 0 or 1. The controlled-not operation on two qubits can be written compactly in the computational basis as
$CNOT(1,2)\ket{x_1}\ket{x_2}=\ket{x_1}\ket{x_2 \oplus x_1}$ where qubit 1 is the control qubit and $\oplus$ represents addition modulo 2. Through out this manuscript, we take advantage of the following relations:
\begin{eqnarray*}
CNOT(j,k)CNOT(j,k)&=&I\\
Z_jZ_j&=&I\\
CNOT(j,k)Z_jCNOT(j,k)&=&Z_j\\
CNOT(j,k)Z_kCNOT(j,k)&=&Z_jZ_k\\
CNOT(j,k)Z_jZ_kCNOT(j,k)&=&Z_k.
\end{eqnarray*}

The last two relationships convert between Ising couplings, $Z_jZ_k$, and local magnetic fields, $Z_k$. The repeated application of CNOT transformations is an explicit method to obtain the zero-field partition function for any Ising tree Hamiltonian of $N$ spins, which is always equivalent to a single free spin and $N-1$ independent spins in a magnetic field \cite{Eggarter1974}. The same transformation applied to trees that are graphs with a few cycles results in partition functions of clusters of spins. All of these partition functions are products of partition functions of few spins and do not lead to any singularities of the zero-field thermodynamic response functions. They do, however, lead to differences in magnetic behavior.

\section{Analytical Solution for the Partition Function}
The stabilizer is defined as all the products of Pauli operators that act trivially on the code space. For the bit-flip code, we can choose any set of pairwise Ising interactions that generates the stabilizer operators. These generators form a Hamiltonian that is similar to the ferromagnetic Ising model,
\begin{equation}
H=-J\sum_{\langle i,j \rangle}Z_{i}Z_{j},
\label{eq:IsingHamiltonian}
\end{equation}

\noindent where $\langle i,j \rangle$ indicates a sum over nearest neighbors, and $J$ sets the energy scale of the problem with temperature measured in units of \Tunit. The choice of generators determines the structure and properties of the system \cite{Viteri2009}.

\subsubsection{\label{sec:cnot_Isingtrees}Ising trees}
A tree is a connected graph without cycles or loops. As a result, there is one and only one path between any two nodes. In an Ising tree, the nodes represent bits and the edges represent the Ising interaction. Each node, $n$, is connected to a single parent, $n_p$, and one or more children, $n_c$. If the node $n$ is at a distance $d$ from the root, the parent is at a distance $d-1$, and the children are at a distance $d+1$. For convenience, we define a function $D$ that converts labels to the minimum distances from the root, $e.g.$, if $D(n)=d$ then $D(n_c)=d+1$. We label each node by its number and its parent's number to make explicit the tree nature of the graph. The Hamiltonian for $N$ spins is then written as
\begin{equation}
H=-J\sum_{n=0}^{N-1}\sum_{n_c} Z_{[n_p,n]}Z_{[n,n_c]},
\label{eq:parchildH}
\end{equation}

\noindent and the Z operator on the root is labeled $Z_{[0,0]}$ although the root has no parent.

The $CNOT([n_p,n],[n,n_c])$ operator transforms $Z_{[n_p,n]}Z_{[n,n_c]}$ into $Z_{[n,n_c]}$, but also transforms $Z_{[n,n_c]}Z_{[n_c,n_{gc}]}$ into $Z_{[n_p,n]}Z_{[n,n_c]}Z_{[n_c,n_{gc}]}$, where $gc$ labels the children of the children. By applying CNOTS first at the outermost connections (leaves) and then moving inwards, we can effectively transform all of the Ising terms into single spin terms.

We define
\begin{equation}
U=\prod_{d=0}^{d_{max}-1}\prod_{D(n)=d}CNOT([n_p,n],[n,n_c])
\label{eq:transformationU}
\end{equation}

\noindent and the product implies right multiplication. Applying this unitary to the Hamiltonian of Eq.~\ref{eq:parchildH} yields
\begin{eqnarray}
H^\prime&=&UHU^\dagger \nonumber\\
&=&-J\sum_{n=1}^{N} Z_{[n_p,n]},
\label{eq:H1freebasis}
\end{eqnarray}

\noindent which represents $N-1$ spins in a magnetic field and one free spin. We will refer to this basis as the {\em free-spin basis} and the original computational basis as the {\em Ising basis}. $U$ is the transformation matrix between the two bases (see Fig.~\ref{fig:structures}).

In Ising trees, every qubit, except the root, has the Hamiltonian $H_1=-JZ$ in the free-spin basis. The partition function is then simply the product of the partition function of a single spin in a magnetic field: $Q_1=\exp{(J/k_BT)}+\exp{(-J/k_BT)}$. The thermodynamic density matrix for a single spin is $\rho=1/2 [I_{2\times 2} + \tanh(J/k_BT) Z]$. The density matrix is used to calculate the polarization in the free-spin basis, $\epsilon=\rm{Tr}\left[Z\rho\right]=\tanh(\it{J/k_BT})$, and the internal energy, $<E_1>=-J\rm{Tr}[Z\rho]=-J\epsilon$. The total thermal density matrix for all of the $N$ spins is a tensor product over independent spin density matrices,
\begin{equation}
\rho_{total}=\otimes_{n=0}^{N-1} \rho_{[n_d,n]} = I_{2\times 2}\otimes_{n=1}^{N-1} \rho,
\label{eq:totalrho}
\end{equation}

\noindent and the total internal energy is then $<E_{total}>=Tr[H'\rho_{total}]=-J(N-1)\epsilon$.

\subsubsection{\label{sec:cnot_decotrees}Ising trees with cycles}

The CNOT Ising tree transformation can also be applied to graphs that can be decomposed into a spanning tree and Ising couplings between spins with the same parent (siblings). The Hamiltonian for $N$ spins is now
\begin{equation}
H=-J\sum_{n=0}^{N-1}\sum_{n_c} Z_{[n_p,n]}Z_{[n,n_c]} -J \sum_{\substack{\langle n,m\rangle \\n_p=m_p}} Z_{[n_p,n]}Z_{[n_p,m]}
\end{equation}
and the same unitary of Eq. \ref{eq:transformationU} transforms it to the free-spin basis, thus
\begin{equation}
H^\prime=-J\sum_{n=1}^{N} Z_{[n_p,n]}-J \sum_{\substack{\langle n,m\rangle \\n_p=m_p}} Z_{[n_p,n]}Z_{[n_p,m]}.
\label{eq:H23freebasis}
\end{equation}
\noindent This is the Hamiltonian of one free spin and finite Ising graphs of sibling spins in non-zero magnetic field. 

Here we examine connections only between sibling pairs, that is the triangular cycles in Structures 2, 3, and 4 (Fig.~\ref{fig:structures}). In this case, there are three types of spins: i) the root which is depolarized in the free-spin basis and has $\left<E_0\right>=0$, ii) the spin that is not connected to a sibling and is described by $H_1=-JZ$, which is equivalent to a spin in a magnetic field, and iii) spins that are connected to a sibling that have the two-spin Hamiltonian $H_2=-J(Z_i+Z_j+Z_iZ_j)$. The expected energy of the spins with the magnetic field Hamiltonian is $\left<E_1\right>=-J\epsilon$ with magnetization $\epsilon=\tanh(J/k_BT)$. The partition function of the siblings is $Q_2=\exp{(3J/k_BT)}+3\exp{(-J/k_BT)}$, and the two-spin density matrix is then $\rho_{i,j}=1/4(I_{4x4}+\alpha Z_i+\alpha Z_j+\alpha Z_iZ_j)$, where 
\begin{equation}
\alpha=\frac{\exp{(3J/k_BT)}-\exp{(-J/k_BT)}}{ \exp{(3J/k_BT)}+3\exp{(-J/k_BT)}}.
\end{equation}
\noindent The energy is $<E_2>=-3J\alpha$, and the magnetization of a single spin is $\alpha$.

The total internal energy is the sum of energies for the three types of spin, $<E_{total}>=<E_1>N_{1}+<E_2>N_{2}/2=-J(\epsilon N_{1}+\frac{3}{2}\alpha N_{2})$. The internal energies for Structure 1, Structure 2, Structure 3, and Structure 4 are then $-J\epsilon(3^k-1)$, $-J[\epsilon(3^{k-1}-1)+\frac{3}{2}\alpha(2\cdot 3^{k-1})]$, $-J\frac{3}{2}\alpha(3^k-1)$, and $-J[\epsilon(2\cdot 3^{k-1})+\frac{3}{2}\alpha(3^{k-1}-1)]$, respectively.

The expectation value of the operators constructed from products of $Z$'s can be calculated quickly from the density matrices for the three spin types. The root is unpolarized, $\rho_0=1/2 I_{2x2}$, and as a consequence any operator that contains $Z_{[0,0]}$ will be zero. The single spins will contribute $\epsilon$ per $Z$. The paired spins are correlated and will contribute $\alpha$ for individual Z's ($Z_i$,~$Z_j$) or the product ($Z_iZ_j$). These rules are sufficient to calculate the magnetic properties of the system and have a succinct description in terms of the geometry. 

\subsection{\label{sec:magnetization} Calculation of the magnetization and the magnetic susceptibility}
The magnetization operator in the computational or Ising basis is $M=\sum_{n= 0}^{N-1} Z_{[n_p,n]}$ and its expectation value is zero by symmetry. The product of the magnetization of each spin and the magnetization of the root define the relative magnetization operator
\begin{equation}
\tilde{M}=\sum_{n=0}^{N-1} Z_{[0,0]}Z_{[n_p,n]}.
\label{eq:mag}
\end{equation}

\noindent The root spin is sufficient to bias the system into one of the two states that break the symmetry~\cite{Chandler_book}, and the $\left<\tilde{M}\right>/N$ is non-zero in the thermodynamic limit when the system is in a ferromagnetic phase. The square of the magnetization relates to the magnetic susceptibility per spin as follows:
\begin{eqnarray}
\chi = \frac{\left<\tilde{M}^2\right>-\left<\tilde{M}\right>^2}{Nk_BT} = \frac{\left<M^2\right>-\left<\tilde{M}\right>^2}{Nk_BT}
\label{eq:magsus}
\end{eqnarray}

Each of the $Z_{n_p,n}$ operators must be transformed into the free-spin basis in order to calculate the magnetic properties. The basis transformation of Eq. \ref{eq:transformationU} maps each $Z_{n_p,n}$ operator onto a product of $Z$'s. When $n$ is at a distance $d$ from the root, the transformation yields
\begin{equation}
UZ_{[n_{d-1},n]}U^\dagger=Z_{[n_{d-1},n]}Z_{[n_{d-2},n_{d-1}]}...Z_{[0,n_1]}Z_{[0,0]},
\label{eq:UZU}
\end{equation}

\noindent with each parent labeled as $n_{d-1}$. The operator $Z_{[n_p,n]}$ becomes a product of $Z$'s on every node on the path from the root to the spin $n$.

The key observation is that the local magnetization operators in the Ising basis are transformed into paths in the free-spin basis. Calculations can then be performed using the paths as follows: 
\begin{itemize}
\item{Label the edges of the tree-like graph with paired siblings by $\alpha$, if the edge is part of a triangle, or by $\epsilon$, otherwise.}
\item{Define $\mathrm{Path}(n,l)$ as the product of the edge labels between nodes $n$ and $l$ along the shortest path.} 
\end{itemize}

\begin{figure}
\centering
\includegraphics[width=0.45\textwidth]{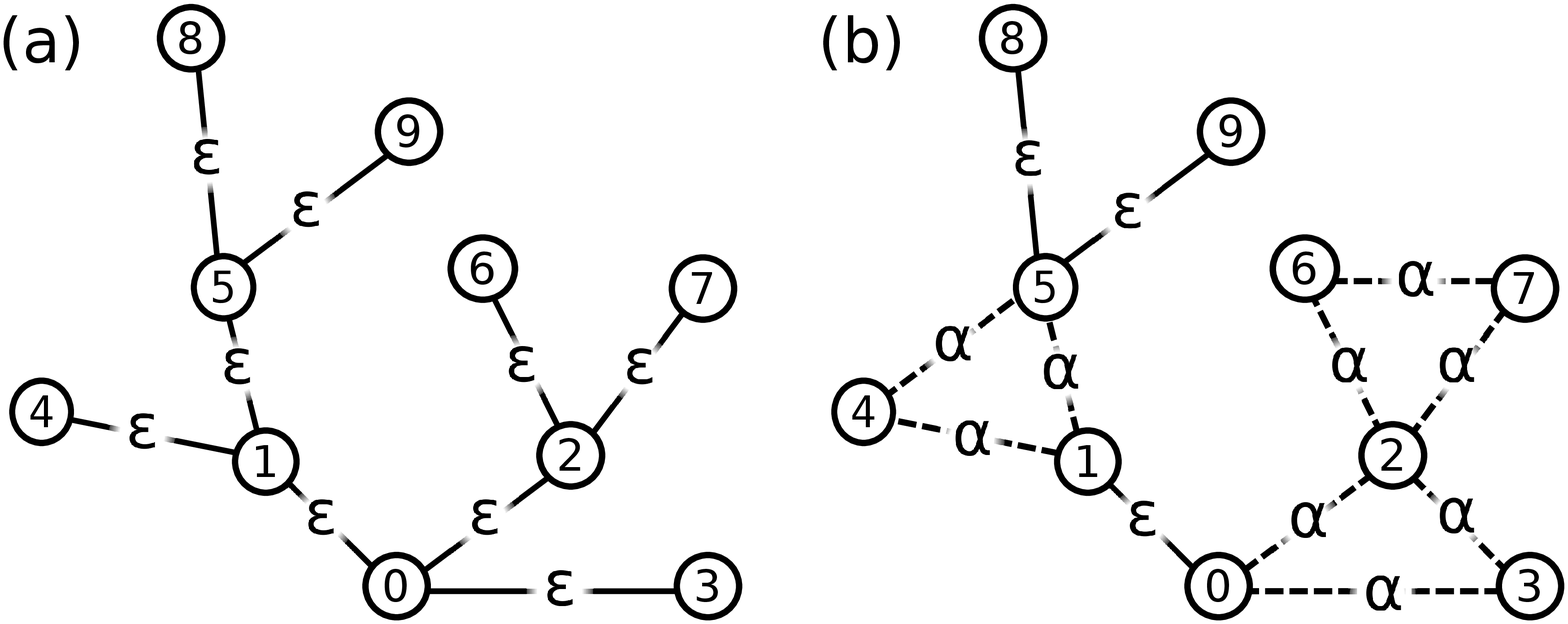}
\caption{\label{fig:extree} Example of labeling edges in (a) trees and (b) tree-like graphs to calculate the magnetic properties based on paths between nodes (see text).}
\end{figure}

\noindent Fig. \ref{fig:extree} shows a tree and related tree-like graph with the edges labeled. As an example we calculate $\mathrm{Path}(4,9)$. For the tree (Fig. \ref{fig:extree}a), $\mathrm{Path}(4,9)=\mathrm{Path}(4,1)\mathrm{Path}(1,5)\mathrm{Path}(5,9)=\epsilon^3$. In the tree-like graph (Fig. \ref{fig:extree}b), there is a {\em shortcut} between the paired sibling nodes 4 and 5 and $\mathrm{Path}(4,9)=\mathrm{Path}(4,5)\mathrm{Path}(5,9)=\epsilon\alpha$.

As shown in Appendix \ref{App:Trees}, the magnetic thermodynamic averages can be related to the paths as 
\begin{equation}
\langle \tilde{M} \rangle= \sum_{n=0}^{N-1} \mathrm{Path}(0,n) \label{eq:pathM}
\end{equation}
and
\begin{equation}
\langle M^2 \rangle =\sum_{n=0}^{N-1}\sum_{l=0}^{N-1} \mathrm{Path}(n,l).
\end{equation}

\noindent For trees, these expressions simplify to
\begin{equation}
\langle \tilde{M} \rangle= \sum_{d=0}^{d_{max}} f(d)\epsilon^d
\label{eq:IsingtreeM}
\end{equation}
and
\begin{equation}
\langle M^2 \rangle =N+\sum_{d=1}^{2d_{max}} 2\phi(d)\epsilon^d,
\label{eq:IsingtreeM2}
\end{equation}
\noindent where $f(d)$ is the number of nodes a distance $d$ from the root and $\phi(d)$ is the number of unidirectional paths of length $d$.

In summary, notice that in the Ising basis, $H$ encodes the geometry by selecting which spins are paired (Eq.~\ref{eq:IsingHamiltonian}), and that the magnetization operator is independent of the connectivity of the $N$ spins. In the free-spin basis, $H'$ is independent of the graph for trees with $N$ nodes (Eq.~\ref{eq:H1freebasis}), and the geometry is now encoded in the magnetization operator (Eq.~\ref{eq:pathM}).

This is well illustrated by calculating the magnetization for two simple examples: a line of $N$-spins and $N-1$ spins connected to a central spin. In both cases the transformation to the free-spin basis results in a Hamiltonian of $N-1$ spins in a magnetic field and a single free spin. As a result, the partition function and density matrix in the free-spin basis are equivalent; however, the magnetizations are quite different. For a line, $f(d)=1$, therefore $\left<\tilde{M}\right>=\sum_{d=0}^{N-1}\epsilon^d$ converges to $1/(1-\epsilon)$ in the limit of large $N$. This yields the familiar result that the magnetization per spin is vanishingly small for $T>0$. For the central spin case, $d=1$ or $0$ and $f(1)=N-1$, with the resulting magnetization being $\left<\tilde{M}\right>=(N-1)\epsilon+1$. The system has non-zero magnetization per spin for all $T<\infty$.

Below we use the equations derived in this section to find analytical expressions for the magnetization and the susceptibility of the stabilizer structures of Fig.~\ref{fig:structures}. All of these systems grow in size as $N=3^k$ as they are based on the concatenation of three units of $3^{k-1}$ spins at each level $k$. The path from the root to the furthermost spin is of length $d_{max}=k$.

\subsubsection{\label{sec:calc_structure1}Structure 1}
In the Ising tree labeled Structure 1 of size $N=3^k$, the number of nodes at distance $d$ from the root is
\begin{equation}
f(d,k)=2^d{k \choose d},
\label{eq:fdk}
\end{equation}

\noindent and according to Eq.~\ref{eq:IsingtreeM}, the expected relative magnetization of Structure 1 at level $k$ is then
\begin{equation}
\left<\tilde{M}(k)\right>_{S1}=\sum_{d=0}^{k}2^d{k \choose d}\epsilon^d=(1+2\epsilon)^k.
\label{eq:S1M}
\end{equation}

\noindent One can understand the result by imagining building up the tree level-by-level. The level $k$ adds 2 nodes to every node in a level $k-1$ tree. The paths between nodes and the root in the inner $k-1$ tree are the same, and the leaves add two paths that are one edge longer. This results in the recursion formula:
\begin{equation}
\left<\tilde{M}(k)\right>_{S1}=(1+2\epsilon)\left<\tilde{M}(k-1)\right>_{S1}.
\end{equation} 

\noindent Notice that this last equation also generates Eq. \ref{eq:S1M}, thus the relative magnetization per spin at zero magnetic field is
\begin{equation}
m_{0S1}(k)=\frac{\left<\tilde{M}_{S1}(k)\right>}{N}=\left(\frac{1+2\epsilon}{3}\right)^k,
\end{equation}

\noindent which vanishes in the limit of large $k$ for all $\epsilon<1$ and $T>0$. In order to calculate the magnetic susceptibility using Eq.~\ref{eq:magsus}, we need to first evaluate the magnetization squared operator. Starting from $k-1$, two leaves are added to every node. Each path of length $d>0$ on the $k-1$ tree now has two extra leaves on each end. This results in one path of length $d$, four paths of length $d+1$, and four paths of length $d+2$. For the $3^{k-1}$ paths of $d=0$, there are now two paths of length one, one path of length two, and two new paths of zero length. Using these observations and defining $\phi_1(d,k)$ as the number of paths of distance $d$ between two spins, Eq.~\ref{eq:IsingtreeM2} can be written recursively:
\begin{eqnarray}
\nonumber
\left<M^2(k)\right>_{S1}&=&3^k+\sum_{d=1}^{2k}2\phi_1(d,k) \nonumber \\
	&=&3^{k-1}\left(1+2+4\epsilon +2\epsilon^2\right) \nonumber \\
	& & +\left(1+4\epsilon+4\epsilon^2\right)\sum_{d=1}^{2(k-1)}2\phi_1(d,k-1) \nonumber \\
	&=&(1+2\epsilon)^2\left<M^2(k-1)\right>_{S1} \nonumber \\
	& &+2(1-\epsilon^2)3^{k-1}.
\end{eqnarray}
\noindent The solution to the recursion formula is 
\begin{widetext}
\begin{equation}
\left<M^2(k)\right>_{S1}= (1+2\epsilon)^{2k}+2(1-\epsilon^2)(1+2\epsilon)^{2(k-1)}
\left\lbrace\frac{1-[3/(1+2\epsilon)^2]^k}{1-3/(1+2\epsilon)^2}\right\rbrace
\label{eq:S1M2}
\end{equation}
\end{widetext}
and the magnetic susceptibility per spin is then
\begin{equation}
\chi_{S1}(k)=\frac{2(1-\epsilon^2)(1+2\epsilon)^{2(k-1)}\{1-[3/(1+2\epsilon)^2]^k\}}{Nk_BT\left[1-3/(1+2\epsilon)^2 \right]}.
\label{eq:S1chi}
\end{equation}

\subsubsection{\label{sec:calc_structure2}Structure 2}
Structure 2 is similar to Structure 1 but the leaves are connected forming triangular cycles. The number of spins at the minimum distance $d$ from the root is the same as in Structure 1 but now there are single spins and spin pairs in the free-spin basis. For paths that include leaf spins from Structure 1, the magnetization needs to include the polarization of a spin pair, $\alpha$. Structure 2 with $3^k$ spins is equivalent to Structure 1 with $3^{k-1}$ spins with a sibling pair connected to each spin. The magnetization is then
\begin{equation}
\left<\tilde{M}(k)\right>_{S2}=(1+2\alpha)\left<\tilde{M}_{S1}(k-1)\right>.
\label{eq:S2M}
\end{equation}

The thermodynamic average of $\left<M^2\right>$ for a Structure 2 of $3^k$ nodes can be built from a Structure 1 with $3^{k-1}$ nodes by examining the extra shortest paths due to the attached outer cycles. The main difference is that the two new nodes connected to the $k-1$ structure are a distance 1 apart instead of a distance 2. The result is that
\begin{eqnarray}
\left<M^2(k)\right>_{S2}&=&\left(1+2\alpha\right)^2 \left<M^2(k-1)\right>_{S1} \nonumber \\
& &+2\cdot 3^{k-1}\left(1+\alpha-2\alpha^2\right)
\label{eq:S2M2}
\end{eqnarray}
and
\begin{equation}
\chi_{S2}(k)=\frac{(1+2\alpha)^2\chi_{S1}(k-1)+2\left(1+\alpha-2\alpha^2\right)/k_BT}{3}.
\label{eq:S2chi}
\end{equation}

\subsubsection{\label{sec:calc_structure3}Structure 3}
In this structure, each spin is part of a triangular cycle, and all spins but the root are paired spins. The thermodynamic average of the magnetization is identical to Structure 1 except the polarization is now $\alpha$ instead of $\epsilon$.
\begin{equation}
\left<\tilde{M}(k)\right>_{S3}=(1+2\alpha)^k.
\label{eq:S3M}
\end{equation}

The magnetization squared depends on the number of shortest paths between all spins, which is quite different from Structure 1 due to shortcuts made by triangular cycles. Applying the same building method of adding nodes to the core yields the following recursion relation
\begin{eqnarray}
\left<M^2(k)\right>_{S3}&=&(1+2\alpha)^2\left<M^2(k-1)\right>_{S3} \nonumber \\
& &+2(1+\alpha-2\alpha^2)3^{k-1},
\end{eqnarray}
\noindent whose solution is:
\begin{widetext}
\begin{equation}
\left<M^2(k)\right>_{S3}=(1+2\alpha)^{2k}+2(1+\alpha-2\alpha^2)(1+2\alpha)^{2(k-1)}
\left\lbrace\frac{1-[3/(1+2\alpha)^2]^k}{1-3/(1+2\alpha)^2}\right\rbrace.
\label{eq:S3M2}
\end{equation}
\end{widetext}
The magnetic susceptibility is then
\begin{eqnarray}
\chi_{S3}(k)&=&\frac{2(1+\alpha-2\alpha^2)(1+2\alpha)^{2(k-1)}}{Nk_BT} \nonumber \\
& &\times \frac{\{1-[3/(1+2\alpha)^2]^k\}}{\left[1-3/(1+2\alpha)^2\right]}.
\label{eq:S3chi}
\end{eqnarray}

\subsubsection{\label{sec:calc_structure4}Structure 4}
In Structure 4, each of the spins form part of triangular cycles except the outer nodes. The relationship between Structure 4 and Structure 3 is similar to the relationship between Structure 2 and Structure 1, and the magnetization properties are calculated to be
\begin{equation}
\left<\tilde{M}(k)\right>_{S4}=\left<\tilde{M}(k-1)\right>_{S3}(1+2\epsilon),
\label{eq:S4M}
\end{equation}
\begin{eqnarray}
\left<M^2(k)\right>_{S4}&=&\left(1+2\epsilon\right)^2 \left<M^2(k-1)\right>_{S3}\nonumber \\
& &+2\cdot 3^{k-1}\left(1-\epsilon^2\right),
\label{eq:S4M2}
\end{eqnarray}
and
\begin{eqnarray}
\chi_{S4}(k)=\frac{(1+2\epsilon)^2\chi_{S3}(k-1)+2\left(1-\epsilon^2\right)/k_BT}{3}.
\label{eq:S4chi}
\end{eqnarray}

\subsection{\label{sec:extension} Extension to the Canonical Stabilizers}
The stabilizer formalism of quantum computing defines a subspace of $n$-qubits by a set of commuting observables that are products of Pauli matrices on the $n$-qubits and have the value of 1 on the subspace. The stabilizer generators are independent operators, trace orthogonal, and commute with one another. As a result, there is always a unitary transformation which maps the stabilizer elements to $Z$ operators on independent spins. Furthermore, this unitary can be constructed from CNOTs, Hadamards, and Pauli matrices \cite{Nielsen_book}.

Structure 1 is derived from the three-qubit classical stabilizer code. The choice of generators is chosen to form an Ising tree, and this is not the standard choice. The standard choice is to use logical Ising interactions at every level of encoding. This choice results in generators that are multi-qubit interactions which grow exponentially with the level of encoding.

\begin{figure}
\centering
\includegraphics[width=0.45\textwidth]{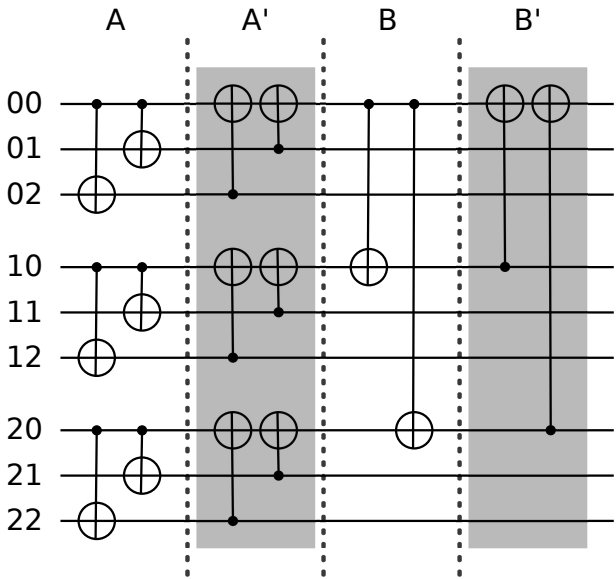}
\caption{\label{fig:circuit} A description of $U$ as a quantum computing circuit for 9 qubits. For the tree and tree-like structures examined, the CNOTs are applied with the control towards the root starting from the leaves and then moving down layers until the root (A,B). For the full-stabilizer, the direction of control is alternated before applying the CNOTs at the next layer (A, A$^\prime$, B, B$^\prime$). }
\end{figure}

The transformation that takes the stabilizer elements to independent Z's is closely related to the transformation used for Structures 1, 2, 3 and 4. Instead of simply applying the CNOTs with the control on the inner node and then progressing inward, the control is alternated from inner to outer. A comparison of the two transformations is shown for 9 qubits in Fig. \ref{fig:circuit}. For Structures 1, 2, 3, and 4 only A and B are applied. For the full stabilizer, A, A$^\prime$, B, and B$^\prime$ are all applied. The detailed description of the transformation can be found in the Appendix \ref{App:Can}. The expected relative magnetization and the magnetization squared are respectively:
\begin{equation}
\left<\tilde{M} (k) \right> = 1 + 2\epsilon\left\lbrace\frac{1-[(2+\epsilon)\epsilon^3]^k}{1-(2+\epsilon)\epsilon^3}\right\rbrace
\label{eq:FSM}
\end{equation}
and
\begin{equation}
\left<\tilde{M^2} (k) \right> = 3^k + 2 \cdot 3^{k-1} \zeta \left[ \frac{1-(\zeta^2/3)^k}{1-(\zeta^2/3)} \right ],
\label{eq:FSM2}
\end{equation}
where $\zeta=(2+\epsilon)\epsilon$.

\section{\label{sec:ResDis}Results and Discussion}
\subsection{\label{sec:appTc}Apparent Critical Temperature for Finite Size Systems}
\begin{figure}
\centering
\includegraphics[width=0.45\textwidth]{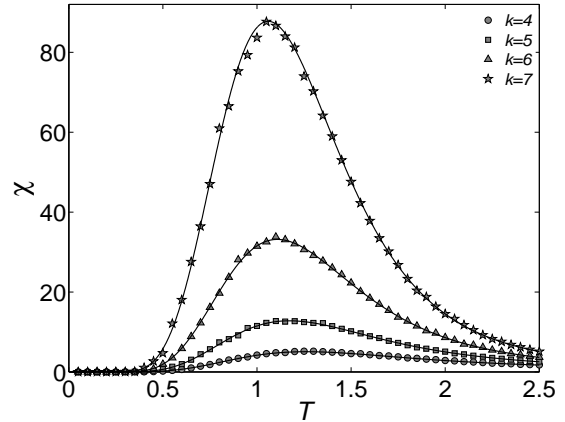}
\caption{\label{fig:numan} Magnetic susceptibilities per spin as a function of temperature $T$ in units of \Tunit\ for different concatenation levels of Structure 3. The solid lines are calculated from Eq.~\ref{eq:S3chi} and the symbols are the result of Monte Carlo simulations~\cite{Viteri2009} using $5 \times 10^{4}$ independent spin configurations.} 
\end{figure}

The analytical results obtained from Eqs.~\ref{eq:S1chi}, \ref{eq:S2chi} and \ref{eq:S3chi} match perfectly with our previous Monte Carlo simulations~\cite{Viteri2009}. As an example, Fig.~\ref{fig:numan} compares the closed form equation of the magnetic susceptibility for Structure 3 with numerical simulations of systems of various sizes. The susceptibilities are calculated using the thermodynamic statistics of $5 \times 10^{4}$ independent spin configurations generated with Wolff cluster simulations at different temperatures.

\begin{figure}
\centering
\includegraphics[width=0.45\textwidth]{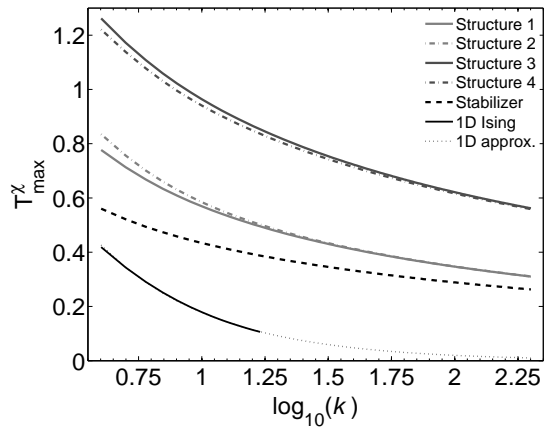}
\caption{\label{fig:Tchimax} Temperature of maximum relative magnetic susceptibility \Tchimax\ for different memory stabilizers of sizes that span from tens to $10^{95}$ spins. Note that 1.70 and 2.23 in the abscissa correspond respectively to the Avogadro's number and to the predicted number of hadrons in the observable universe~\cite{Stosic2005}.}
\end{figure}

For finite-size self-correcting memories, the susceptibility as a function of temperature shows a maximum which occurs at an apparent critical temperature \TchimaxN. It is clear from Fig.~\ref{fig:numan} that this temperature decreases with system size as expected. Finite size effects replace the divergences at the thermodynamic critical point by finite peaks shifted away from \Tc~\cite{Fisher1972, Newman_book}. Previously~\cite{Viteri2009}, we used a first order approximation to estimate these shifts for the case of susceptibility. A fit of \Tchimax\ against the system size $N$ gave us an estimate for \Tc, \cio\, and \nup. With only four data points, we forecasted a finite \Tc. Analytical solutions for the magnetic susceptibility permit the study of bigger systems and present a more complete picture of the finite-size effects on the magnetic properties. We obtain \TchimaxN\ numerically and Fig.~\ref{fig:Tchimax} compares them for the four Ising stabilizer structures, the canonical stabilizer, and the 1D Ising model as a function of total number of spins in a double log scale, $\rm{log}_{10}(\rm{log}_{3}(N))$. The numerical calculation of \Tchimax\ for the 1D Ising model of systems bigger than $3^{18}$ spins results in a numeric underflow. The dotted line in the figure is an extrapolation using a two parameters fit of the solid line to the equation \Tchimax$=ak^{-b}$.

In the limit of systems of infinite size, Eqs.~\ref{eq:S1M}, \ref{eq:S2M}, \ref{eq:S3M}, \ref{eq:S4M}, and \ref{eq:FSM} reveal that the only temperature at which \mo\ takes the exact value of one is \Tc$=0$.  
It is seen in Fig.~\ref{fig:Tchimax} that \Tchimax\ converges very slowly to zero with system size. Based on the closed form equations this prolonged decay cannot be captured by a simple first order equation in $N^{-1/\nu'}$ nor by any finite power expansion in $N$ without including an offset. We cannot calculate numerically the size of memory stabilizers with a \Tchimax\ of practically zero before we run into numerical overflow. As shown in the figure, memory stabilizers utilizing all the observable matter in the universe will still behave as a finite-size system with almost all of their spins correlated at a finite apparent critical temperature on the order of \Tunit. This is in contrast to the linear spin case where \Tchimax\ rapidly approaches zero with increasing system size. For 3 spins, Structure 1 and the line are equivalent with \Tchimax$=1.07$. We can then ask how many spins are required to reach a certain \Tchimax. As an example, a maximum susceptibility of \Tchimax$=0.29$ is achieved using $3^6$ spins in a line, but it would require a Structure 1 of $3^{313}$ spins. This presents an interesting challenge as these networked spin systems stand in contrast to our standard notion of what size the thermodynamic limit is appropriate.

The thermal stability of the information encoded into finite systems for the four structures and the full-stabilizer can be related to \Tchimax\ (Fig.~\ref{fig:Tchimax}). The choice of stabilizers leads to a significant change in this apparent critical temperature. Two thirds of the spins in Structure 2 form closed cycles, and the other third of spins form a core that is the same as a $k-1$ Structure 1. As expected for small systems, Structure 2 remains magnetized for a broader range of temperatures than Structure 1. Similarly, Structure 4, with 2/3 of its spins as free leaves, is less stable to temperature driven fluctuations than Structure 3. The cores of Structures 1 and 2, and Structures 3 and 4, account for 1/3 of the spins, and they show similar magnetic behavior in the astronomical limit of large $k$. The canonical choice of stabilizer elements remains magnetized below a broad range of temperatures, and shows the same long ranged order properties, but under this thermodynamic criteria, is a less efficient memory stabilizer than the simpler pairwise interaction geometries.

\subsection{\label{sec:pwrlaws}Power-law Correlations and Finite Size Effects}
Assuming that there is a relation between a 1-D path of correlated spins and the total size of the system, $L=N^{1/d}$, we can define a correlation length exponent scaled to the system size $\nu'=\nu\cdot d$. The dimension, $d$, of each of the structures of Fig.~\ref{fig:structures} is unknown. According to the standard scaling hypothesis, and provided that the system size is large enough, the following scaling properties are expected at the critical point: $c(N)\propto N^{\frac{\alpha}{\nu'}}$, $m(N)\propto N^{-\frac{\beta}{\nu'}}$, and $\chi(N)\propto N^{\frac{\gamma}{\nu'}}$~\cite{Newman_book}. Closed form equations for the magnetization and the magnetic susceptibility can be used to calculate critical exponents.

An analytical formula for the relative magnetization critical exponent can be obtained by equating the relative magnetization per spin to the $N^{-\frac{\beta}{\nu'}}$ power law. After simplification of the exponent $k$, the \betnu\ critical exponent can be written in terms of the logarithm of the relative magnetization as follows:
\begin{equation}
\beta/\nu' = 1-\frac{\rm{ln}(\psi)}{\rm{ln}(3)},
\label{eq:betanu}
\end{equation}
\noindent where $\ln(\psi)=\ln(\langle\tilde{M}\rangle)/k$. For Structures 1 and 3, $\psi$ is independent of $k$ and we find $\psi_1=1+2\epsilon$ and $\psi_3=1+2\alpha$, respectively. For Structures 2 and 4, $\psi$ depends weakly on $k$ as $\psi_2=\psi_1\left(\frac{\psi_3}{\psi_1}\right)^{1/k}$ and $\psi_4=\psi_3\left(\frac{\psi_1}{\psi_3}\right)^{1/k}$. The power law for the relative magnetization per spin can thus be written as a function of any arbitrary temperature:
\begin{equation}
m_0(T)=N^{-\left [1-\frac{\ln \left (\psi(T) \right )}{\ln(3)}\right]}.
\label{eq:magfuncT}
\end{equation}
This last equation reduces the magnetic susceptibility of Eqs. \ref{eq:S1chi} and \ref{eq:S3chi} to the form:
\begin{equation}
\chi(k,T)=A(k,T)\cdot (3^k)^{-2\beta/\nu' +1},
\label{eq:chipwrlw}
\end{equation}

\noindent where for Structures 1 and 3 $A$ is
\begin{equation}
A(k,T)_{S1}=\frac{(1-\epsilon^2)[(1+2\epsilon)^{2k}-3^k]}{k_BT(2\epsilon^2+2\epsilon-1)(1+2\epsilon)^{2k}}
\label{eq:AS1}
\end{equation}
\noindent and
\begin{equation}
A(k,T)_{S3}=\frac{(1+\alpha-2\alpha^2)[(1+2\alpha)^{2k}-3^k]}{k_BT(2\alpha^2+2\alpha-1)(1+2\alpha)^{2k}},
\label{eq:AS3}
\end{equation}

\noindent respectively. As $T$ approaches the thermodynamic critical temperature of $T_c=0$, the $A$ function of Eqs. \ref{eq:AS1} and \ref{eq:AS3} is independent of the system size and converges respectively to
\begin{equation}
A(T)_{S1}\simeq\frac{(1-\epsilon^2)}{k_BT(2\epsilon^2+2\epsilon-1)}
\label{eq:AS1ap}
\end{equation}
\noindent and
\begin{equation}
A(T)_{S3}\simeq\frac{(1+\alpha-2\alpha^2)}{k_BT(2\alpha^2+2\alpha-1)}.
\label{eq:AS3ap}
\end{equation}

\begin{figure}
\centering
\includegraphics[width=0.45\textwidth]{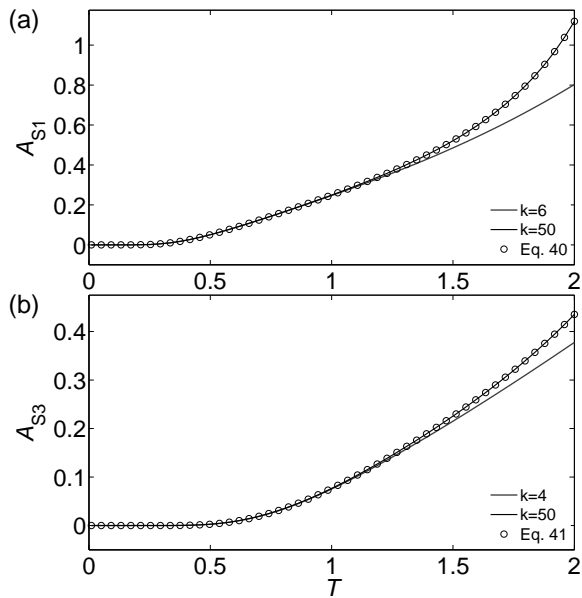}
\caption{\label{fig:pwrlaws} Power law proportionality function $A(k,T)$ for (a) Structure 1 and (b) Structure 3. Solid lines are calculated using Eqs. \ref{eq:AS1} and \ref{eq:AS3} while the circles come from the approximations of Eqs. \ref{eq:AS1ap} and \ref{eq:AS3ap}. In a broad region of temperatures near the thermodynamic critical point, these simple equations are enough to define a magnetization power law on $N$ for any system size.}
\end{figure}

In this limit, the magnetic susceptibility can be written as the well known formula $\chi(N)=A(T)\cdot N^{\frac{\gamma}{\nu'}}$. Examining Eq. \ref{eq:chipwrlw}, we find that $\gamma/\nu' + 2\beta/\nu' = 1$, which matches the Rushbrooke and Josephson scaling law $d = \gamma/\nu + 2\beta/\nu$ if written as a function of the correlation size exponent \nup. Fig. \ref{fig:pwrlaws} tests the temperature region in which this approximation holds for all system sizes. There is a broad temperature region where the magnetic properties are well described by $N$ and temperature dependent critical exponents.

The set of critical exponents obtained numerically for particular ill-predicted critical temperatures (see Tables IV and V of Ref.~\cite{Viteri2009}) match very well with those calculated using Eq.~\ref{eq:betanu} and the hyperscaling relation. These apparent critical temperatures fall in the temperature region in which Eqs.~\ref{eq:AS1ap} and \ref{eq:AS3ap} hold.

We find that there is a broad temperature region above \Tc\ where the \betnu\ exponent is almost zero (it is strictly zero only at $T=0$). Complementary, and for the same broad temperature region, the \gamnu\ critical exponent reaches almost the value of one. The interpretation is simple: the number of correlated spins grows almost at the same rate as $N$. Spins that present long ranged correlations develop a net macroscopic alignment when an infinitesimal magnetic field is applied~\cite{Chandler_book}.

\section{\label{sec:Con}Conclusion}
We use CNOT gates to transform the Hamiltonian of stabilizer structures into a Hamiltonian consisting of uncoupled single and pairs of spins. In the original basis, the Hamiltonian encodes the geometry by selecting which bits interact. The magnetization operator is independent of the graph for $N$ spins. In the free spin basis, the Hamiltonian is independent of the graph for $N$ spins and the geometry is now encoded in the magnetization operator. This transformation allows us to obtain an analytical partition function and closed form equations for the effective magnetization and susceptibility with respect to a central spin.

The analytical solutions match very well with the numerical results presented previously\cite{Viteri2009} for finite size systems of $N \leq 3^8$ spins. With a slight modification of the transformation sequence we calculate the analytical magnetization and susceptibility of the canonical stabilizer Hamiltonian.

In our previous calculation based on four values of $k$, we forecast a finite critical temperature for systems of infinite size. Our analytical solution shows that this prediction is incorrect. After applying a sequence of CNOT operations on the four stabilizer Hamiltonians studied in this work, the partition function results in a collection of free elements. The interactions represented in the magnetization operator yield to graphs that have a transition from magnetic to random at \Tc=0 in the thermodynamic limit. However, they possess unusual long-range-order properties as previously observed in hierarchical systems which also have \Tc=0 (e.g.: Sierpinski gaskets~\cite{Liu1985} and Cayley trees~\cite{Stosic2005}). The memory stabilizer structures develop spontaneous magnetization below an apparent critical temperature for unrealistically large systems. The relative magnetization persists below a finite temperature for systems of $N=3^{200}$ spins. For a practical implementation, the infinite system is a poor approximation, and it remains poor even for finite systems that are astronomical in size. This conflicts with our notion of the thermodynamic limit, where the infinite system well describes crystalline solids of few billion unit cells~\cite{Liu1985}.

First order finite-size scaling analysis is incomplete and fails to describe the slow decrease of \Tchimax\ with the system size for all structures. For systems with no phase transition at finite temperatures, the shift away from \Tc\ cannot be written as a simple power expansion in $N^{1/\nu'}$. The partition function of the systems, and its first and second derivatives with respect to an external magnetic field, do not present critical points. They are continuous, well behaved, and show spontaneous magnetization for a broad range of temperatures. In the broad region near \Tc=0, scaling properties of the magnetization and magnetic susceptibility satisfy power-laws as a function of $N$. 

The memory stabilizers presented in this work do not show a phase transition in the thermodynamic sense. However, for a wide range of temperatures and finite size, there are many long paths of correlated spins that go through the structure resulting in a net macroscopic magnetization. These structures have free energy functions that spontaneously break global symmetry in the presence of environment-induced fluctuations, thus they stabilize the memory.

We arrive to similar conclusions as in our previous work~\cite{Viteri2009}. Fig.~\ref{fig:Tchimax} suggests that one way to increase the apparent critical temperature for a system of a given finite-size is by adding generators to each spin site. Structure 3 is the best self-correcting memory as it has the broadest range of temperatures in which the system remains magnetized. The four simple two-body-interaction structures investigated have different levels of connectivity. We find that the relationship between coordination number and the apparent finite size critical temperature \Tchimax\ is not obvious. The number of generators is less important than the structure. The canonical stabilizer Hamiltonian seems to be thermodynamically a less stable memory than the simpler pairwise based construction with the minimum number of generators (Structure 1), but it is more stable than the Ising chain. Kinetically, the canonical stabilizer could be the the most impervious to fluctuations. For systems with all of the spins aligned, the lowest excited state energy for Structure 1 is $2J$ from the ground state, but this gap grows as $2kJ$ for the canonical stabilizer. The multi-body interactions of the canonical stabilizer result in many large kinetic barriers that may be advantageous for preserving certain spin configurations.

Finally, the exploration of stabilizer Hamiltonians defined by geometries of non-integer dimensions could yield self-correcting quantum memories with few multi-qubit interactions. Small finite size systems show an unusual order preservation for a broad range of temperatures making them suitable for a practical implementation of passive error correction.

\begin{acknowledgments}
Authors thank Keisuke Fujii for pointing out that it is possible to obtain an analytical partition function for Structure 1 by applying a CNOT transformation. This work was supported by Georgia Institute of Technology.
\end{acknowledgments}

\bibliography{Stab-basis-PM}% Produces the bibliography via BibTeX.

\appendix
\section{Trees and trees with cycles}
\label{App:Trees}

For an Ising tree, there is only one path between any two nodes $n$ and $l$. We define $P_{l,n}$ as the product of Z operators of all of the nodes on the path between $n$ and $l$ on the spanning tree (including $n$ and $l$). The transformation of Eq. \ref{eq:UZU} can then be written succinctly as
\begin{equation}
UZ_{[n_p,n]}U^\dagger=P_{0,n}
\end{equation}

\noindent and $M$ transforms to
\begin{equation}
UMU^\dagger=\sum_n UZ_{[n_p,n]}U^\dagger=\sum_n P_{0,n}.
\end{equation}

\noindent The expected value must be zero by symmetry and this is easy to confirm since each term contains the root factor $Z_{[0,0]}$. Applying the transformation to the $\tilde{M}$ operator of Eq.~\ref{eq:mag}, results in
\begin{eqnarray}
U\tilde{M}U^\dagger&=&\sum_n UZ_{[n_p,n]}Z_{[0,0]}U^\dagger \nonumber\\
&=&\sum_n P_{0,n}P_{0,0} \nonumber\\
&=&\sum_n Z_{[n_{d-1},n]}Z_{[n_{d-2},n_{d-1}]}...Z_{[0,n_1]}Z_{[0,0]}Z_{[0,0]} \nonumber\\
&=&\sum_n Z_{[n_{d-1},n]}...Z_{[0,n_1]} \nonumber\\
&=&\sum_n P_{n,n_1}. \label{eq:magAppA}
\end{eqnarray}

\noindent Note that none of the terms contain $Z_{[0,0]}$ and the number of $Z$ factors in each term is the distance between $n$ and the root.

The expectation value of the magnetization is the product of the polarization of all the spins on the path from $n$ to $n_1$ in the free-spin basis. For the structures studied in Section~\ref{sec:magnetization}, the spins that are in sibling pairs have polarization $\alpha$ and otherwise have polarization $\epsilon$ (except the root). This leads the relative magnetization to be
\begin{equation}
\left<\tilde{M}\right>=\sum_{n}\alpha^{c_n}\epsilon^{D(n)-c_n},
\label{eq:expM}
\end{equation}

\noindent where $c_n$ is the number of spins that are in a pair between the root and $n$. One can express this graphically by labeling every edge in the graph with an $\epsilon$ if it is not part of a triangle, and with an $\alpha$ if it is part of a triangle (see Fig.~\ref{fig:extree}). One then starts from a node and multiplies the label of the edges between the node and the root on the spanning tree. Summing over all nodes yields Eq.~\ref{eq:pathM}, and a comparison to Eq.~\ref{eq:magAppA} shows that ${\mathrm{Path}}(0,n)=\left<P_{n,n_1}\right>$. 

The magnetization operator squared, $M^2=\tilde{M^2}$, in the transformed basis is
\begin{eqnarray}
UM^2U^\dagger&=&\sum_{n,l} UZ_{[l_p,l]}Z_{[n_p,n]}U^\dagger\nonumber\\
&=&\sum_{n,l} P_{0,l}P_{0,n}.
\label{eq:UM2U}
\end{eqnarray}

\noindent The product of $P_{0,l}$ and $P_{0,n}$ results in the $Z$'s that are in the intersection of the paths from the root to $l$, and from the root to $n$, to cancel. The $Z$ operator on the last node in common is $Z_{{\mathrm{Last}}(l,n)}$, and this node is included in the path from $l$ to $n$ on the spanning tree. As an example, consider node 4 and node 9 of Fig. \ref{fig:extree}. Then $P_{0,4}=Z_{[1,4]}Z_{[0,1]}Z_{[0,0]}$, $P_{0,9}=Z_{[5,9]}Z_{[1,5]}Z_{[0,1]}Z_{[0,0]}$, and $P_{4,9}=Z_{[1,4]}Z_{[1,0]}Z_{[1,5]} Z_{[5,9]}$. Node 1 is the last node in common and, as a result, $P_{0,4}P_{0,9}=P_{4,9}Z_{[0,1]}$ and $Z_{[0,1]}=Z_{{\mathrm{Last}}(4,9)}$. These operators are the same for Fig. \ref{fig:extree}(a) and (b), but the expectation values differ as explained below. 

Eq.~\ref{eq:UM2U} is simplified to:
\begin{eqnarray}
UM^2U^\dagger&=&\sum_{n,l} P_{l,n}Z_{{\mathrm{Last}}(l,n)} .
\label{eq:UM2Usimp}
\end{eqnarray}

\noindent To calculate the expectation value of $P_{l,n}Z_{n,l}$, we must consider three cases. In the first case, the node $n$ is contained in the path between the root and $l$ or vice-versa, and $Z_{{\mathrm{Last}}(l,n)}$ equals $Z_{[n_p,n]}$ or $Z_{[l_p,l]}$, respectively. The expectation value of $P_{l,n}Z_{n,l}$ is simply the polarization of spins on the path from $l$ to $n$ excluding the node ${\mathrm{Last}}(l,n)$. The polarization of each node is the label of the edge connecting it to its parent and as a result $\langle P_{n,l}Z_{n,l}\rangle={\mathrm{Path}}(n,l)$.
In the second case, the two nodes after the last node are not part of the same triangle. The polarization of spins at these nodes are independent and again $\langle P_{n,l}Z_{n,l}\rangle={\mathrm{Path}}(n,l)$. In the third case, the two nodes after the last node are part of the same triangle. The polarizations are not independent and two $Z$'s yield a single $\alpha$. This is equivalent to taking a {\em shortcut}, and for all cases, $\langle P_{n,l}Z_{n,l}\rangle={\mathrm{Path}}(n,l)$. Combining these observations with Eq. \ref{eq:UM2Usimp}, we obtain
\begin{equation}
\left<M^2\right>=\sum_{n,l} \mathrm{Path}(n,l).
\label{eq:expM2}
\end{equation}

\section{Canonical Stabilizer}
\label{App:Can}

To define the full stabilizer, it is useful to start at the top level and work down. At level $k$ there is one qubit (labeled 0), composed of three level $k-1$ qubits labeled 00, 01, and 02. We can define the logical operator as

\begin{equation}
Z^{(k)}_0=Z^{(k-1)}_{00}Z^{(k-1)}_{01}Z^{(k-1)}_{02}
\end{equation}
and the highest order stabilizer elements as
\begin{eqnarray}
A^{(k)}_{01}&=&Z^{(k-1)}_{00}Z^{(k-1)}_{01}\nonumber\\
A^{(k)}_{02}&=&Z^{(k-1)}_{00}Z^{(k-1)}_{02}.
\end{eqnarray}

\noindent Continuing this procedure, we define
\begin{eqnarray}
Z^{(j)}_{\eta}&=&Z^{(j-1)}_{\eta 0}Z^{(j-1)}_{\eta 1}Z^{(j-1)}_{\eta 2}\nonumber\\
A^{(j)}_{\eta 1}&=&Z^{(j-1)}_{\eta 0}Z^{(j-1)}_{\eta 1}\nonumber\\
A^{(j)}_{\eta 2}&=&Z^{(j-1)}_{\eta 0}Z^{(j-1)}_{\eta 2}
\end{eqnarray}
and one composite operator,
\begin{eqnarray}
A^{(j)}_{\eta 0}=A^{(j)}_{\eta 1}A^{(j)}_{\eta 2},
\end{eqnarray}
where $\eta$ is a $k-j+1$ string of trits. It is convenient to consider $\eta$ as a number in base 3. Also, we introduce four useful identities:
\begin{eqnarray}
A^{(j)}_{\eta 1}Z^{(j)}_{\eta}&=&Z^{(j-1)}_{\eta 2}\label{eq:A1Z2}\\
A^{(j)}_{\eta 2}Z^{(j)}_{\eta}&=&Z^{(j-1)}_{\eta 1}\label{eq:A2Z1}\\
A^{(j)}_{\eta 0}Z^{(j)}_{\eta}&=&Z^{(j-1)}_{\eta 0}\label{eq:A0Z0}\\
A^{(j)}_{\eta t} A^{(j)}_{\eta t}&=&I\label{eq:AA}.
\end{eqnarray}
Note how the products of $A^{j-1}$ with $Z^{j}$ interchange the 1 and 2 labels. 

The Hamiltonian is 
\begin{equation}
H=\sum_{j=1}^k\sum_{x=1}^2\sum_{\eta=0}^{3^{k-j}-1} A^{j}_{\eta x},
\end{equation}
and there exists a unitary that transforms $A$'s to single qubit $Z$'s. The chosen unitary performs the following transformation:
\begin{eqnarray}
UA^j_{\eta x}U^\dagger&=&Z_{\eta x 0^{(j-1)}} \nonumber \\
UA^j_{\eta 0}U^\dagger&=&UA^j_{\eta 1}U^\dagger UA^j_{\eta 2}U^\dagger \nonumber \\
&=&Z_{\eta 1 0^{(j-1)}} Z_{\eta 2 0^{(j-1)}}, 
\end{eqnarray}
where $0^{l}$ is a string of $l$ zeros and $x=1$ or $2$. Every physical qubit is denoted by a ${k+1}$ trit string with the first trit set to zero. This will transform the $3^k-1$ stabilizer elements into $3^k-1$ single $Z$ operators. To further specify the unitary, we set
\begin{equation}
UZ^k_0U^\dagger=Z_{0^{k+1}}.
\end{equation}

An explicit construction of $U$ is as follows. Arranging the spins on the tree that defines Structure 1 and starting at the spins at the maximum distance from the root, $d_{max}$, apply $CNOT(n_{d_{max}-1},n_{d_{max}})$ between all spins connected on the graph at this distance. Then apply $CNOT(n_{d_{max}},n_{d_{max}-1})$ to the same spins. Then move up one level and repeat the procedure first applying $CNOT(n_{d_{max}-2},n_{d_{max}-1})$ at this distance and then $CNOT(n_{d_{max}-1},n_{d_{max}-2})$. Continue to the root. If we only used the CNOTs that are controlled by parents, we have the same unitary as Structure 1. Switching to CNOTs that are controlled by children allows us to transform logical Z's into Z's on single spins. The unitary is shown as a circuit for $k=2$ in Fig. \ref{fig:circuit}. 

$U^\dagger Z_{\mu}U$ must be determined to calculate the magnetization.
This is possible by constructing $Z_{\mu}$ from stabilizer elements and $Z^k_0$ using Eqs.~\ref{eq:A1Z2},~\ref{eq:A2Z1}, and~\ref{eq:A0Z0}. If we write $\mu=\mu_k\mu_{k-1}...\mu_1\mu_0$, one can show that 
\begin{equation}\label{eq:Zmu}
Z_{\mu}=Z^k_0\prod_{j=1}^{k}A^j_{\mu_k\mu_{k-1}...\bar{\mu}_{j-1}},
\end{equation}
where
\begin{equation}
\bar{\mu}_j= (2\mu_j)\mod 3
\end{equation}
which exchanges 1's for 2's and follows from Eqs.~\ref{eq:A1Z2} and \ref{eq:A2Z1}.

$Z^k_0$ is the product of $Z$ on all spins, and the element $Z^k_0A^k_{0x}$ is the product of all spins on a single $k-1$ level qubit. Geometrically, each $A^j$ splits off one block of $3^{j-1}$ spins from a block of $3^{j}$ level. Each $A$ reduces the total number of spin operators by $1/3$ until we reach a single qubit operator. 

To calculate the magnetic properties, we examine the transformed product of two $Z$ operators, $Z_\mu$ and $Z_\nu$. Assume that $\mu$ and $\nu$ agree in the first $q+1$ trits, the product is then
\begin{eqnarray}
Z_\mu Z_\nu&=& \left(Z^k_0\prod_{j=1}^{k}A^j_{\mu_k\mu_{k-1}...\bar{\mu}_{j-1}}\right)\cdot \left( Z^k_0\prod_{j=1}^{k}A^j_{\nu_k\nu_{k-1}...\bar{\nu}_{j-1}}\right) \nonumber \\
&=&\prod_{j=1}^{k-q}A^j_{\mu_k\mu_{k-1}...\bar{\mu}_{j-1}}A^j_{\nu_k\nu_{k-1}...\bar{\nu}_{j-1}}\
\end{eqnarray} 

After the transformation $UZ_\mu Z_\nu U^\dagger$, each remaining $A^{j}_{\eta\{1,2\}}$ corresponds to a single $Z$ operator and each $A^{j}_{\eta 0}$ corresponds to two $Z$ operators. The $Z$'s will be unique between $\mu$ and $\nu$ except for the only case where $\eta$ agrees, $A^{k-q}_{\mu_k\mu_{k-1}...\bar{\mu}_{k-q-1}}$ and $A^{k-q}_{\nu_k\nu_{k-1}...\bar{\nu}_{k-q-1}}$. The number of independent $Z$'s in $UZ_\mu Z_\nu U^\dagger$ denotes an effective distance between $\mu$ and $\nu$, 
\begin{eqnarray}
\delta(\mu ,\nu)&=&2(k-q-1) +{\mathrm{zeros}}(\mu_{k-q-2}...\mu_{0}) \nonumber \\
& &+{\mathrm{zeros}}(\nu_{k-q-2}...\nu_{0})\nonumber \\
& &+ 2^{\mu_{k-q-1}\cdot\nu_{k-q-1}/2}
\end{eqnarray}
in which $q+1$ is the number of leading trits that agree in $\mu$ and $\nu$, and ${\mathrm{zeros}}(\mu_j...\mu_0)$ counts the number of zeros in the string $\mu_j...\mu_0$. The first trits that do not agree, $\mu_{k-q-1}$ and $\nu_{k-q-1}$, result in 1 or 2 independent $Z$'s depending on whether either trit equals zero. For a given $q$, the minimal effective distance between the spins is $2k-2q-1$.

Calculating the relative magnetization requires computing the effective distances between the root $Z_{0^{k+1}}$ and all other spins $Z_\mu$. Clearly, the number of leading trits that agree are the number of leading zeros in $\mu$, $L(\mu)$. The result is
\begin{eqnarray}
\delta(0^{k+1},\mu) &=&2(k-L(\mu)-1) + (k-L(\mu)-1) \nonumber \\
& &+ {\mathrm{zeros}}(\mu_{k-L(\mu)-2}...\mu_0)+1 \nonumber \\
&=& 3k-4L(\mu)+{\mathrm{zeros}}(\mu)-2.
\end{eqnarray}
This effective distance serves the same role as the distance from the root to the nodes in Structure 1.

\begin{figure}
\centering
\includegraphics[width=0.425\textwidth]{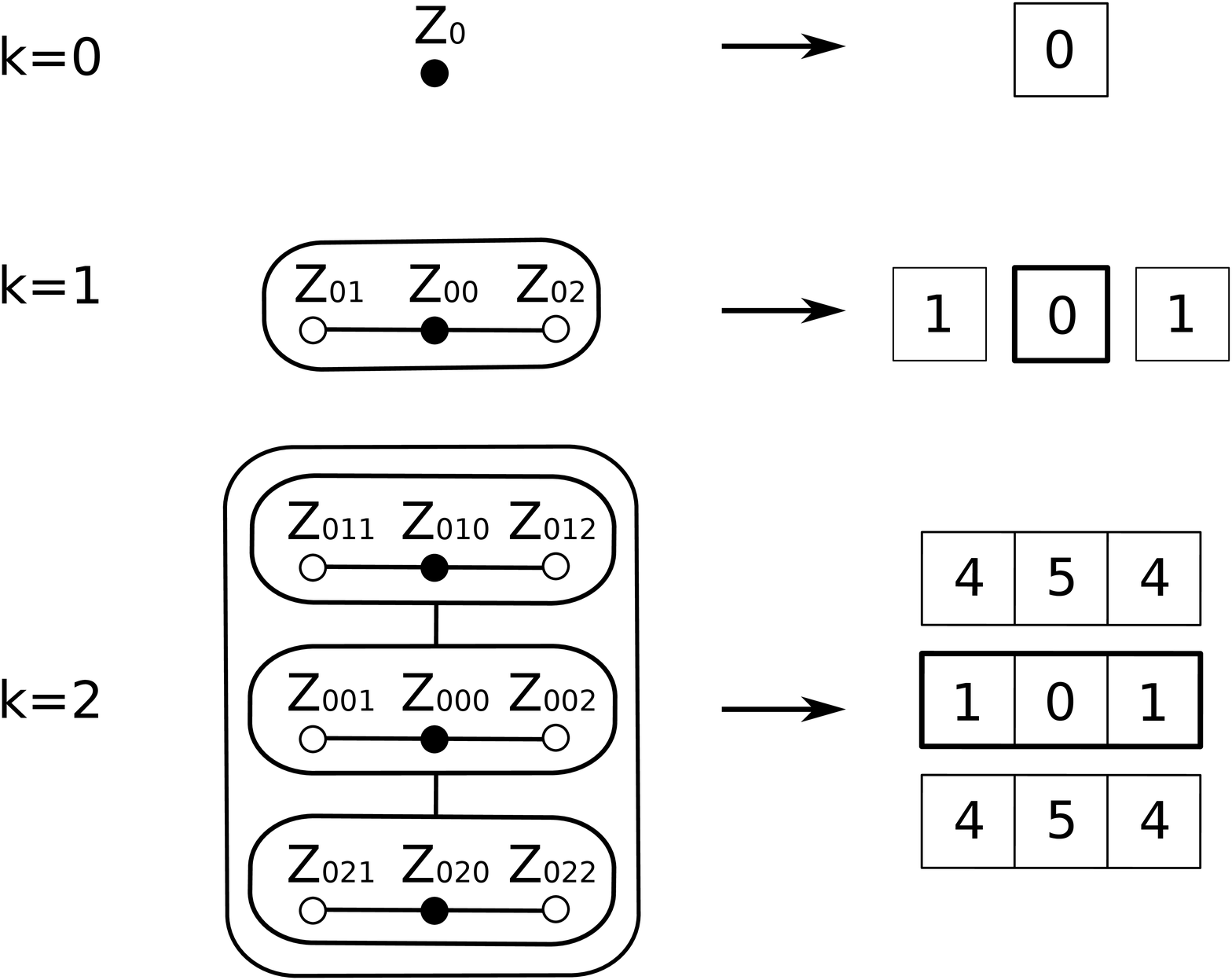}
\caption{\label{fig:stabCounter} The effective distances of the canonical stabilizer up to $k=2$. For each $k$ the black dot represents the root qubit. The corresponding $Z_\mu$ is written next to each qubit. The numbers shown on the right are the effective distances from the root to the qubit at the same position. The bold box contains the effective distances inherited from the level $k-1$.}
\end{figure}

Fig. \ref{fig:stabCounter} shows the effective distance for each spin of the canonical stabilizer up to level $k=2$. Each qubit at level $k$ consists of three qubits of level $k-1$. The effective distances for each level $k$ include effective distances from the central $k-1$ level qubit, and two equivalent sets of distances corresponding to the appended $k-1$ level qubits. The minimum effective distance in each new set is $3k-2$ at the {\em corner} qubit, which occurs when there is exactly one zero in the string $\mu$. In a new set, the effective distance increases by one for each additional zero in $\mu$. Suppose $\mu$ labels a qubit in an appended block, then $\mu_k=0$ and $\mu_{k-1}\neq 0$. If $\mu_j=0$ for $j<{k-1}$, then the qubit is effectively one step farther from the root than if $\mu_j=1$ or 2. Analogous to the Ising trees, the expected value of the magnetization is calculated using the polarization $\epsilon$ and the effective distance of each qubit.

With the minimum effective distance of $3k-2$, the expected value of the magnetization is calculated to be
\begin{eqnarray}
\left<\tilde{M} (k) \right> = \left<\tilde{M} (k-1) \right> + 2 (2 + \epsilon)^{k-1} \epsilon^{3k-2},
\end{eqnarray}
and the solution to this recursion formula is
\begin{eqnarray}
\left<\tilde{M} (k) \right> = 1 + 2\epsilon\left\lbrace\frac{1-[(2+\epsilon)\epsilon^3]^k}{1-(2+\epsilon)\epsilon^3}\right\rbrace.
\end{eqnarray}
The $2+\epsilon$ arises from the three choices of $\mu_j$ and there are $k-1$ values of $j$ in the appended block. 

The thermodynamic average of $\left<M^2(k)\right>$ is obtained by performing a similar calculation but summing over the effective distances between all spins. At level $k$, each level $k-1$ qubit has the same internal magnetization squared, $\left<M^2(k-1)\right>$. To compute the effective distances between qubits in $k-1$ blocks, we determine how many trits are required to distinguish a pair of physical qubits. It takes a single trit to determine which two logical blocks are paired. Each block contains qubits labeled by $k-1$ different trits. Similar to the magnetization, each trit contributes a factor of $2+\epsilon$, but now there are a total of $2k-1$ choices. The minimum effective distance between two sets is also $2k-1$, since only the leading trit can agree (q=0). Overall, the thermodynamic average of $\left<M^2\right>$ is expressed as
\begin{eqnarray}
\left<M^2(k) \right> &=& 3 \left<M^2(k-1) \right>  \nonumber \\
& &+ 2 (2+\epsilon)^{2k-1} \epsilon^{2k-1},
\end{eqnarray}
and the solution to the recursion is given by
\begin{eqnarray}
\left<M^2(k) \right> &=& 3^k + 2 \cdot 3^{k-1} \zeta \sum_{j=0}^{k-1} \frac{\zeta^{2j}}{3^j} \nonumber \\
		&=& 3^k + 2 \cdot 3^{k-1} \zeta \left[ \frac{1-(\zeta^2/3)^k}{1-(\zeta^2/3)} \right ],
\end{eqnarray}
where $\zeta=(2+\epsilon)\epsilon$.

\end{document}